\documentclass[fleqn,usenatbib]{mnras}

\usepackage{CJKutf8}
\usepackage{newtxtext,newtxmath}

\usepackage[T1]{fontenc}

\DeclareRobustCommand{\VAN}[3]{#2}
\let\VANthebibliography\thebibliography
\def\thebibliography{\DeclareRobustCommand{\VAN}[3]{##3}\VANthebibliography}

\usepackage{graphicx}	
\usepackage{amsmath}	
\usepackage{color}
\usepackage{float}
\usepackage{subfigure}
\usepackage{enumerate}
\usepackage{multirow}
\usepackage{booktabs}
\usepackage{threeparttable}
\graphicspath{{./}{figures/}}

\newcommand{\dm}{\,pc\,cm$^{-3}$}
\newcommand{\pu}{J2016$+$3711}
\newcommand{\snr}{CTB\,87}

\newcommand{\ps}{\,{\rm s}^{-1}}
\newcommand{\tc}{\tau_{10}}

\title[Discovery and timing of PSR \pu]{Discovery and timing of pulsar \pu\ in supernova remnant \snr\ with FAST}

\author[Q.C. Liu \& W.J. Zhong et al.]{
Qian-Cheng Liu (\begin{CJK}{UTF8}{bsmi}劉前程\end{CJK}),$^{1,6}$
Wen-Juan Zhong (\begin{CJK}{UTF8}{bsmi}鍾文娟\end{CJK}),$^{1,6}$
Yang Chen (\begin{CJK}{UTF8}{bsmi}陳陽\end{CJK}),$^{1,2}$\thanks{E-mail: ygchen@nju.edu.cn}
Pei Wang (\begin{CJK}{UTF8}{bsmi}王培\end{CJK}), $^{3,4}$\thanks{E-mail: wangpei@nao.cas.cn}
\newauthor{
Ping Zhou (\begin{CJK}{UTF8}{bsmi}周平\end{CJK}),$^{1,2}$
You-Ling Yue (\begin{CJK}{UTF8}{bsmi}岳友嶺\end{CJK}),$^{3,4,5}$
and Di Li (\begin{CJK}{UTF8}{bsmi}李菂\end{CJK})$^{3,4,5}$}{}
\\
$^{1}$School of Astronomy \& Space Science, Nanjing University, Nanjing 210023, China \\
$^{2}$Key Laboratory of Modern Astronomy and Astrophysics,
Nanjing University, Ministry of Education, 
Nanjing 210023, China \\
$^{3}$National Astronomical Observatories, Chinese Academy of Sciences,
Chaoyang District, Datun Road, A.20, Beijing, 100101, China \\
$^{4}$CAS Key Laboratory of FAST, National Astronomical Observatories,
Chinese Academy of Sciences, Beijing 100101, China \\
$^{5}$University of Chinese Academy of Sciences, Beijing 100049, China \\
$^{6}$Contributed equally
}

\date{Accepted 2024 February 01. Received 2024 January 25; in original form 2024 January 04}

\pubyear{2024}

\begin{document}
\label{firstpage}
\pagerange{\pageref{firstpage}--\pageref{lastpage}}
\maketitle

\begin{abstract}
We report on our discovery of the radio pulsar, PSR J2016$+$3711,
in supernova remnant (SNR) \snr, 
with a $\sim 10.8\sigma$ significance of pulses,
which confirms the compact nature of the X-ray point source in \snr.
It is the first pulsar discovered in SNRs using 
Five-hundred-meter Aperture Spherical radio Telescope (FAST).
Its integrated radio pulse profile can be well
described by a single component, with a width 
at 50\% of the peak flux density
of about $28.1^\circ$ 
and an effective width of about $32.2^\circ$.
The mean flux density at 1.25\,GHz is estimated to be about 
$15.5\mu$Jy. Combined with the non-detection of the radio pulse at lower
frequencies, the radio spectral index of the pulsar is constrained 
to be $\lesssim 2.3$.
We also present the timing solution based on 28 follow-up FAST observations.
Our results reveal a period of 50.81 ms, period derivative of $7.2\times 10^{-14}$ s$\ps$, and dispersion measure of 
428 pc\,cm$^{-3}$.
The strength of the equatorial surface magnetic dipole
magnetic field is inferred to be about $1.9\times10^{12}$ G.
Using the ephemeris obtained from the radio observations, we searched \emph{Fermi}-LAT data for gamma-ray pulsations but detected no pulsed signal.
We also searched for radio pulses with FAST toward the X-ray counterpart of the gamma-ray binary HESS J1832$-$093 proximate to SNR G22.7$-$00.2 but found no signal.
\end{abstract}

\begin{keywords}
pulsars: individual: PSR J2016$+$3711 -- ISM: individual objects: CTB 87 (G74.9$+$1.2) -- telescopes.
\end{keywords}



\section{Introduction}\label{sec:introduction}
Pulsar wind nebulae (PWNe), one of the most important components 
of supernova remnants (SNRs), are produced by the relativistic winds of rotation-powered neutron stars as they interact with the surroundings. 
Nonetheless, only about one
percent of the known pulsars are found to be directly associated with 
PWNe and SNRs \citep[e.g.][]{lyn12}.
On the other hand,
although the existence of pulsars in PWNe is beyond doubt, 
about 30\% of the known PWNe and PWN candidates
have not been confirmed with pulsed emission signals
to date (according to \citealt{fer12}).
Finding pulsars in PWNe and SNRs is crucial to studying the pulsar formation and the supernova explosion mechanism, and bridging the gap between the theoretical prediction and the observational results.

SNR \snr, appearing located in a superbubble \citep{liu18} and in physical contact with a molecular cloud \citep[e.g.][]{liu18, kot03},
is characterized by the hosted PWN with trailing morphology in X-rays
\citep{mat13}.
A point-like X-ray source, CXOU J201609.2+371110,
has been found in the southeastern head portion of the trailing structure, which was suggested to be the pulsar candidate \citep{mat13}.
Several surveys have covered SNR \snr\ to search for the radio pulsar.
Using the Low Frequency Array (LOFAR), the point-like X-ray source was observed at about 150\,MHz with a sensitivity of about 0.4\,mJy \citep{str19}.
Using the Arecibo radio telescope, \snr\ was fully searched at 430\,MHz with a sensitivity of 0.05\,mJy \citep{gor96}.
Also, it was searched using the 76\,m Lovell radio telescope at 606\,MHz with sensitivities of 3\,mJy \citep{big96} and 0.16\,mJy \citep{lor98}, respectively. However, no radio pulsation from \snr\ was ever detected in the previous searches.
Proximate to the southwest of SNR G22.7$-$00.2, an X-ray source
(XMMU J183245$-$0921539)
was found consistent with the location of the gamma-ray binary HESS J1832$-$093 \citep{hes15}.
Recently, near-infrared spectroscopic observations found the companion to be an O6 V star \citep{soelen24}, while the nature of the compact object remains unknown.
The X-ray spectrum of XMMU J183245$-$0921539 is well-fitted by a power-law model with a photon index suggesting a pulsar nature \citep{mori17}.
The steep ATCA spectrum of the radio source coincident with the binary is also consistent with the synchrotron radiation from a radio pulsar residing in the binary \citep{tam20}.

Although non-detection of radio pulse may be because the beam of the pulsar does not point toward us, it could also result from the instrumental limitations of the surveys. 
Now, the Five-hundred-meter Aperture Spherical radio Telescope (FAST), with a large effective area, allows us to search for faint pulsars with unprecedented high sensitivity. 
Actually, a number of faint pulsars have been detected \citep[e.g.][]{li18,cam20,zha19,han21} since the first one discovered with FAST \citep{qia19}.

We have observed two X-ray point-like sources, CXOU J201609.2$+$371110 in SNR \snr\ and XMMU J183245$-$0921539 
proximate to
SNR G22.7$-$00.2, respectively, with FAST.
In this paper, we report on our searches for radio pulses
toward them
and the detection of pulses in \snr. 
We describe the observations and reduction of the data in Section~\ref{observation}.
We present our results, the discovery and timing solution of the pulsar in \snr, in Section~\ref{result}. 
We discuss the properties of the detected pulsar in Section~\ref{discussion} and summarise this paper in 
Section~\ref{summary}.

\begin{figure*}
\centering
\includegraphics[width=0.955\columnwidth]{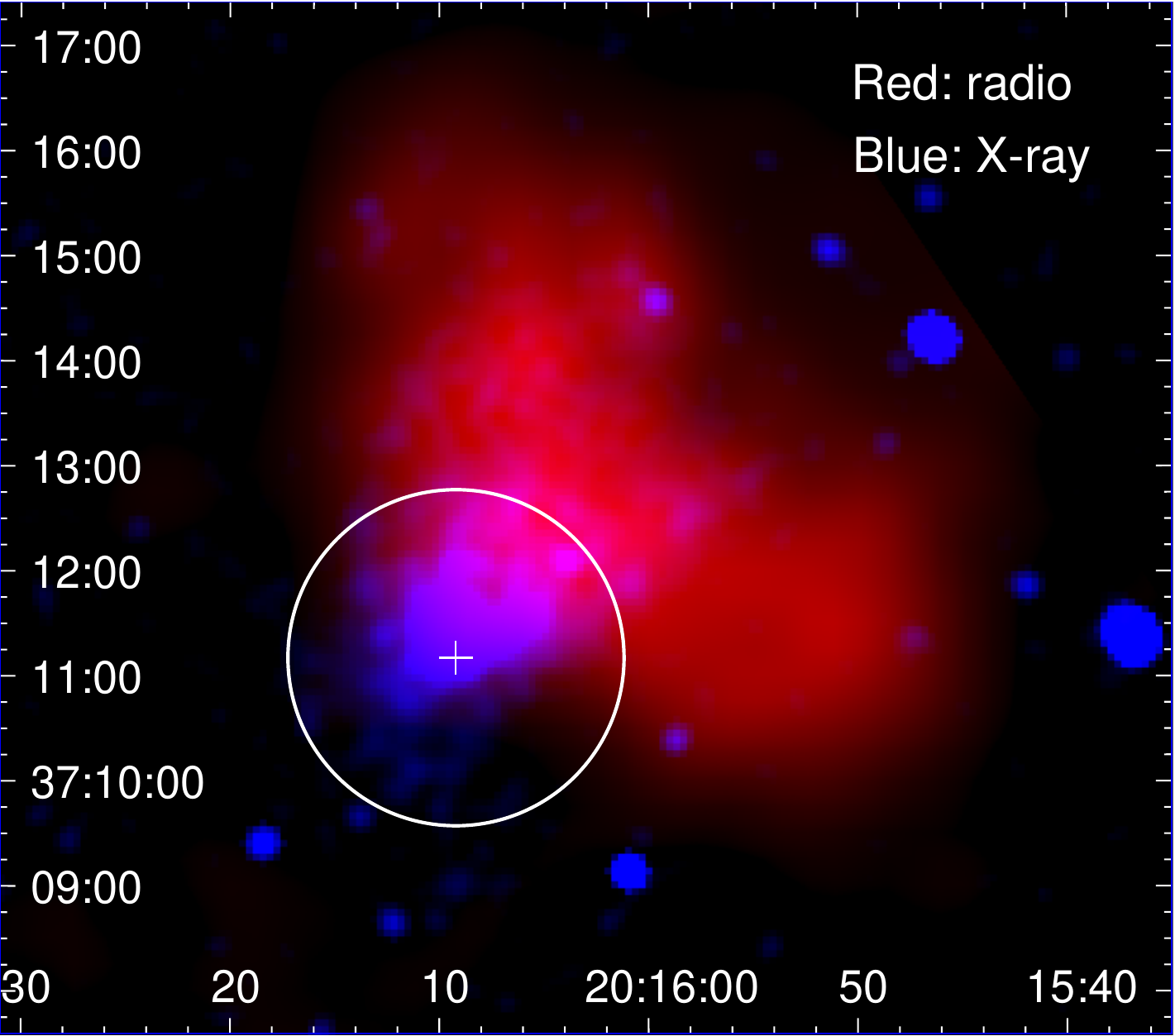}
\includegraphics[width=1.0\columnwidth]{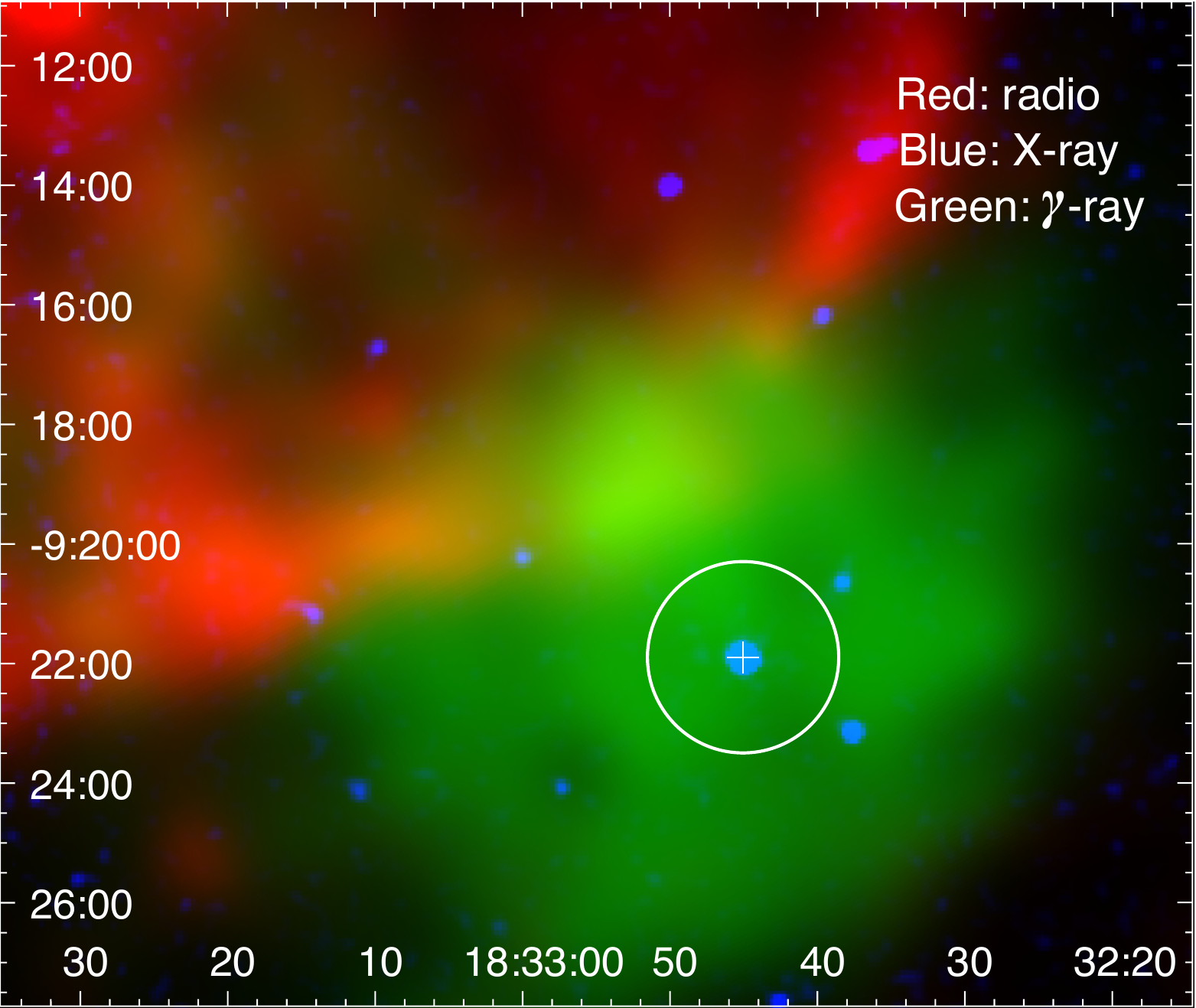}
\caption{(\emph{Left}) Dual-wavelength image of SNR \snr: NVSS 1.4\,GHz 
radio continuum image
in red, and \emph{Chandra} X-ray (0.5--7\,keV) image in blue.
The white cross indicates the location of the X-ray point-like 
source \citep{mat13}. The white circle shows the beam size of FAST at 
1.25\,GHz toward position P$_{\rm CTB87}$.
(\emph{Right}) Tri-wavelength image
of the southwestern part of
SNR G22.7$-$0.02: VGPS 1.4\,GHz 
radio continuum image
in red, the \emph{XMM-Newton} X-ray (0.5--6\,keV) image 
in blue,
and the H.E.S.S.\ gamma-ray ($>$ 1 TeV) integral flux image in green.
The white cross and circle are similar to those in the left panel, but for 
position P$_{\rm G22.7}$.
The radio continuum images in both panels and the gamma-ray image in the right panel are re-gridded with a $4''\times4''$ resolution.}
\label{fig:snr}
\end{figure*}
\begin{table*}
\footnotesize
\caption{Parameters of the targets} \label{tab:para}
\tabcolsep 15pt 
\begin{tabular*}{\textwidth}{lcccccc}
\hline\hline
Source & Position & RA & Dec & Distance $d$ & Predicted DM & DMs searched \\
 & & & & (kpc) & (pc cm$^{-3}$) & (pc cm$^{-3}$) \\
\hline
CTB\,87 & P$_{\rm CTB87}$ & 20:16:09.2 & $+$37:11:10.5 &
$6.1\pm 0.9\,^{a}$ & 197.00$^{+72}_{-98}$ & 50--500 \\
G22.7$-$00.2 & P$_{\rm G22.7}$ & 18:32:45.0 & $-$9:21:53.9 &
$\sim 4.5\,^{b}$ & 471.48 & 150--1000 \\
\hline
\end{tabular*}
    \begin{tablenotes}
        \footnotesize
        \item $^{a}$\citealt{kot03}\\
        \item $^{b}$\citealt{su14}
    \end{tablenotes}
\end{table*}

\section{Observation and Data Reduction}\label{observation}

Our radio observations 
(PI: Y.\ Chen) were carried out with FAST during 2019 July 15--16 in the ``risk-sharing'' open session, for a total time of about 1\,hr for each source.
The two target positions are centred at ($20^{\rm h}16^{\rm m}09^{\rm s}.2$, $+37^\circ11'10''.5$, J2000) 
and ($18^{\rm h}32^{\rm m}45^{\rm s}.0$, $-9^\circ21'53''.9$, J2000) 
(denoted as positions P$_{\rm CTB87}$ and 
P$_{\rm G22.7}$ hereafter) for the point-like X-ray sources
in SNR \snr\ and close to SNR G22.7$-$00.2, respectively
(see Fig. \ref{fig:snr}; \citealt{mat13, hes15}).
In the observations,
the central beam of the 19-beam 1.05--1.45\,GHz receiver \citep{jia19}
was used, with the tracking mode applied. 
The half-power beamwidth of the telescope is about $3.2'$ at 1.25\,GHz and the pointing accuracy of the telescope is better than $16''$.
The digital back-end provided a bandwidth of 400\,MHz with 4096 frequency channels and a time resolution of 49.152\,$\mu$s.
The data were recorded in the search mode of PSRFITS format.

We used the PulsaR Exploration and Search TOolkit ({\small PRESTO}) software
\citep{ran01} to reduce the data. 
The data were first cleaned from the radio frequency interference (RFI) and then were used to search for periodic signals.
The dispersion search ranges of the targets (see Table~\ref{tab:para}) were limited based on the dispersion measure (DM) predicted by the electron density model \citep{yao17}. 
The data were de-dispersed in steps of 0.1\,pc\,cm$^{-3}$ below DM = 230\,pc\,cm$^{-3}$,
steps of 0.3\,pc\,cm$^{-3}$ for 230\,pc\,cm$^{-3} <$ DM $<$ 446\,pc\,cm$^{-3}$,
steps of 0.5\,pc\,cm$^{-3}$ for 446\,pc\,cm$^{-3} <$ DM $<$ 800\,pc\,cm$^{-3}$
and steps of 1\,pc\,cm$^{-3}$ above 800\,pc\,cm$^{-3}$, which were chosen using the \emph{ddplan} routine in {\small PRESTO}. 
We selected candidates of pulse signals with {\small PRESTO}-reported significance above $4\sigma$. 
The selected candidates were then folded and inspected by eyes to clarify them as either RFI or possible radio pulse.
We also searched for single pulses using the \emph{single\_pulse\_search.py} routine in {\small PRESTO}. 
Yet no positive candidate of single pulse has been found toward any of the targets.

After the radio pulses from the \snr\ PWN were discovered in the ``risk-sharing'' observation in 2019 (see \S\ref{result:radio} below), 28 follow-up timing observations were then carried out until 2023, with the same receiver and back-end setup as for the discovery observation.
A summary of the follow-up observations used in the timing analysis is given in Table \ref{tab:tim_obs}.
The pulse time of arrivals (ToAs) was measured for each observation by cross-correlating the pulse profile against a high-S/N template, which was obtained by fitting a set of Gaussian curves to the best detection.
We carried out the subsequent timing analysis of the ToAs with the {\small TEMPO} pulsar timing package
\citep{nice15}\footnote{\url{http://tempo.sourceforge.net}}
to obtain more accurate measurements of the timing parameters.
DM was measured using ToAs from multifrequency subbands.
As shown in Table \ref{tab:tim_obs}, the gap between some observations can be 1 month or even much longer, we thus used the {\small DRACULA} script
\citep{freire18}\footnote{\url{https://github.com/pfreire163/Dracula}}
to help derive phase-connected timing solution for the pulsar.

\begin{figure*}
\centering
\includegraphics[scale=0.6]{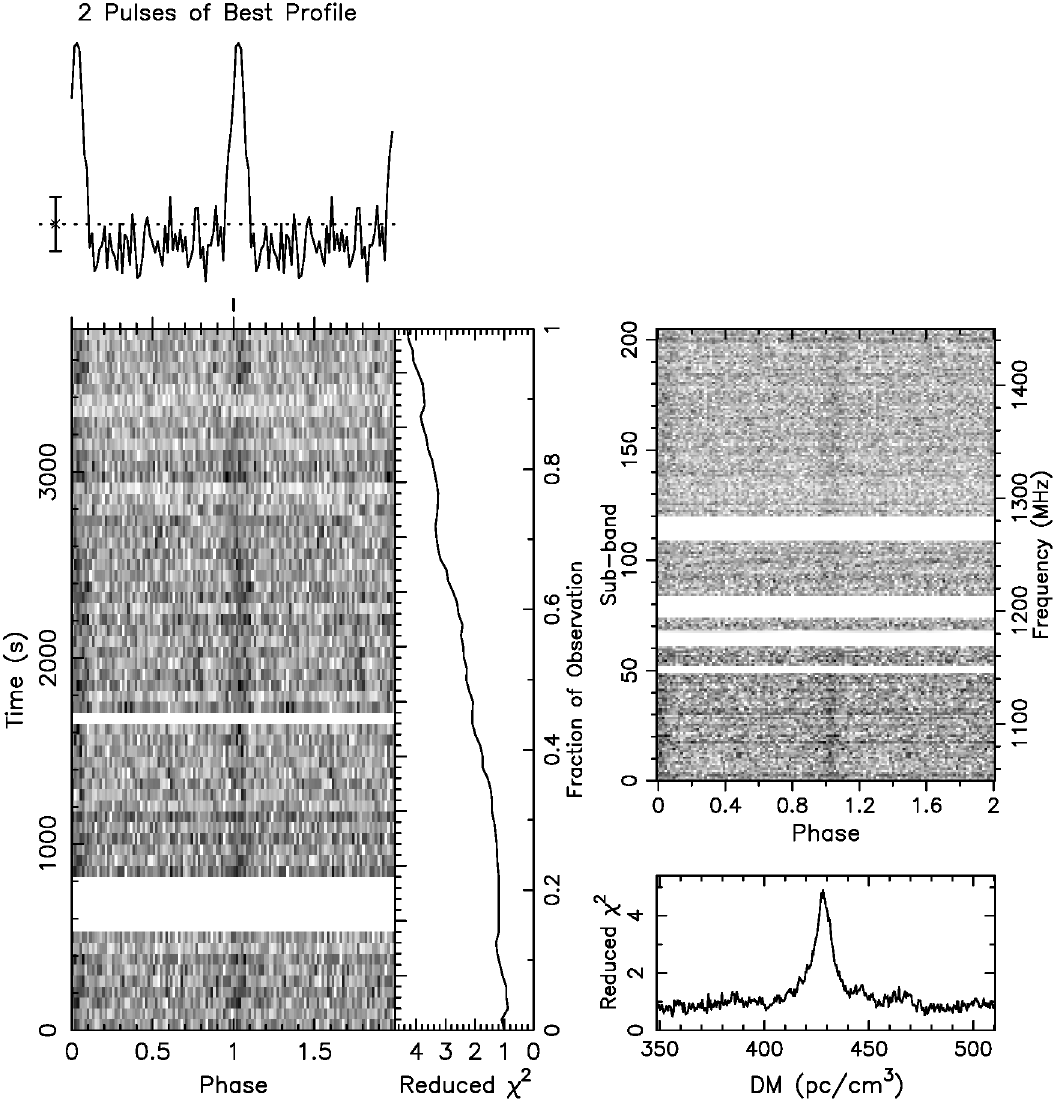}
\caption{Radio pulses detected toward 
position P$_{\rm CTB\,87}$ in
SNR CTB\,87.
The pulsed signal is present throughout the entire 1 hour of the observation 
within the bandwidth of the data (1.05--1.45\,GHz). 
The DM is 429.5(5.0)\dm\ and the
period of the radio pulses is 50.806263(65)\,ms.
}
\label{fig:pulsar}
\end{figure*}

\begin{table}
    \centering
    \caption{Summary of the follow-up observations made with FAST for PSR J2016+3711}
    \setlength{\tabcolsep}{10pt}
    \begin{tabular}{lcccc}
    \hline\hline
      No.& Date & MJD & $T_{\rm obs}$ & Detected $^a$  \\
      & (yyyy-mm-dd) &  & (min) & Y/N \\
    \hline
    1$^*$&	2021 Aug 07&	 59432.8&	60 & Y \\
    2&	2021 Aug 27&	 59452.7&	30 & Y \\
    3&	2021 Aug 29&	 59455.6&	30 & Y \\
    4&	2021 Aug 31&	 59457.5&	30 & Y \\
    5$^*$&	2021 Sep 02&	 59459.5&	30 & Y \\
    6&	2021 Oct 09&	 59496.6&	30 & Y \\
    7&	2021 Oct 16&	 59503.5&	30 & Y \\
    8&	2021 Oct 23&	 59510.5&	30 & Y \\
    9&	2021 Oct 30&	 59517.5&	30 & Y \\
    10$^*$&	2021 Nov 06&	 59524.4&	20 & Y \\
    11$^*$&	2022 Aug 08&	 59799.7&	30 & Y \\
    12&	2022 Oct 03&	 59855.6&	30 & Y \\
    13&	2022 Oct 05&	 59857.6&	30 & Y \\
    14&	2022 Oct 07&	 59859.6&	30 & Y \\
    15&	2022 Oct 09&	 59861.6&	30 & N \\
    16&	2022 Oct 16&	 59868.6&	30 & N \\
    17&	2022 Oct 23&	 59875.6&	30 & Y \\
    18&	2022 Oct 30&	 59882.5&	30 & Y \\
    19&	2022 Nov 06&	 59889.5&	30 & Y \\
    20&	2022 Nov 13&	 59896.5&	30 & Y \\
    21&	2022 Nov 28&	 59911.4&	30 & Y \\
    22$^*$&	2022 Dec 13&	 59926.4&	30 & N \\
    23$^*$&	2022 Dec 28&	 59941.4&	30 & Y \\
    24&	2023 Jan 15&	 59959.3&	30 & Y \\
    25$^*$&	2023 Jan 29&	 59973.3&	30 & Y \\
    26&	2023 Aug 19&	 60175.5&	30 & Y \\
    27$^*$&	2023 Aug 29&	 60184.7&	30 & N \\
    28&	2023 Sep 09&	 60196.6&	30 & Y \\ 
    \hline
    \end{tabular}
    \begin{tablenotes}
        \footnotesize
         \item $^a$ The signal can be seen in every observation by folding the data with the timing solution.
         \item $^*$ The JUMP positions.
    \end{tablenotes} 
    \label{tab:tim_obs}
\end{table}

\section{Results}\label{result}
\subsection{Detection of radio pulses from PSR J2016+3711 and timing}
\label{result:radio}
We did not detect 
radio pulses
toward P$_{\rm G22.7}$ 
\footnote{
According to equation (1) below, the upper limit for the flux-density of P$_{\rm G22.7}$ at 1.25 GHz is roughly estimated to be 5.3 $\mu$Jy for $S_{\rm S/N}=5$ and $W_{\rm eff}=5\%P$ assumed.}
but detected radio pulses toward P$_{\rm CTB87}$ (Fig. \ref{fig:pulsar}), with a significance of $\sim 10.8\sigma$ and a reduced $\chi^2$ of 4.3.
We will refer to this pulsar as PSR~\pu\ hereafter.
The period ($P$) of the detected pulses in the 2019 observation is 50.806263(65)\,ms, and the DM of \pu\ is 429.5(5.0) \dm.
We note that 
the period is somewhat similar to that 
estimated from the X-ray observation of the host PWN
(e.g. $P\approx 80$\,ms in \citealt{mat13}
or 65 ms in \citealt{guest20}).
 
Although unlikely, the radio pulses could potentially arise from
sideband effect of the single-dish telescope. 
To clarify if the detected radio pulses are a result of the effect, we searched for published known pulsars in a region centred at position P$_{\rm CTB87}$ with an angular radius of $5^\circ$ using the ATNF database
\citep{man05}\footnote{\url{https://www.atnf.csiro.au/research/pulsar/psrcat/}},
the Pulsar ALFA Survey Project
\footnote{\url{http://www2.naic.edu/~palfa/newpulsars/index.html}},
the FAST Galactic Plane Pulsar Snapshot survey \citep{han21,zhx23}\footnote{\url{http://zmtt.bao.ac.cn/GPPS/GPPSnewPSR.html}}, 
and the Commensal Radio Astronomy FAST Survey
\footnote{\url{https://crafts.bao.ac.cn/pulsar/}}.
With the DM limited in range 429.5$\pm$100\dm, the results are shown in Table~\ref{tab:knownpulsar}.
There is one pulsar, J2022+3842, which has similar DM and period as those of the radio pulses we detected. We folded our data with the period of J2022$+$3842, yet no radio pulse was detected. 
Therefore, the radio pulsation we detected is from the newly discovered pulsar at position P$_{\rm CTB87}$ in SNR CTB\,87, i.e., \pu.

\begin{figure}
\centering
\includegraphics[width=\columnwidth]{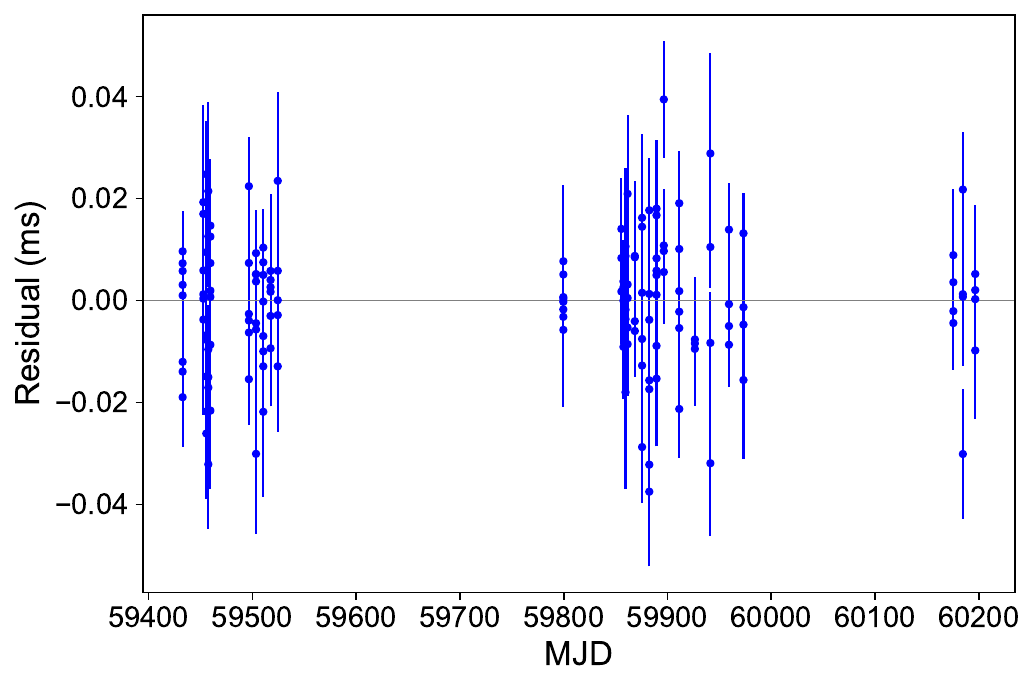}
\caption{Timing residuals from the best-fit timing models presented in Table \ref{tab:tim_results} as a function of the observation date for PSR J2016$+$3711.
}
\label{fig:residual}
\end{figure}

In Table \ref{tab:tim_results}, we present the timing solution for \pu.
We only obtained a timing solution with several JUMPs left (see Table \ref{tab:tim_obs}).
One possible explanation is the discontinuity in the pulsar's frequency or its differential, e.g. glitches, which are commonly observed in young pulsars \citep{lyn12}.
The fitted position of \pu\ is spatially coincident with the X-ray point source detected by \emph{Chandra} \citep{mat13}.
The postfit timing residuals with all of the ToAs obtained by the timing solution are shown in Fig. \ref{fig:residual}.

\begin{table*}
\footnotesize
\caption{Known pulsars near the P$_{\rm CTB87}$ position in 5$^\circ$ region} \label{tab:knownpulsar}
\tabcolsep 25pt 
\begin{tabular*}{\textwidth}{cccccc}
\hline\hline
Pulsar & RA & Dec & Offset & DM & Period \\
 & & & (degree) & (pc cm$^{-3}$) & (ms) \\
\hline
J2004+3304g & 20:04:08 & $+33$:04 & 4.80 & 410.1 & 1193.37  \\
J2004+3429 & 20:04:46.9 & $+34$:29:17.7 & 3.55 & 351.0 & 240.95  \\
J2005+3411g & 20:05:45 & $+34$:11 & 3.77 & 489.0 & 651.05 \\
J2005+3547 & 20:05:17.4 & $+35$:47:25.4 & 2.59 & 401.6 & 615.03 \\
J2005+3552 & 20:05:47.5 & $+35$:52:24.3 & 2.46 & 455.0 & 307.94 \\
J2010+3230 & 20:10:26.5 & $+32$:30:07.3 & 4.83 & 371.8 & 1442.45 \\
J2011+3521g & 20:11:04 & $+35$:20 & 2.12 & 438.1 & 943.23 \\
J2014+3326g & 20:14:24 & $+33$:26 & 3.77 & 332.7 & 977.28 \\
J2021+3651 & 20:21:05.4 & $+36$:51:04.8 & 1.04 & 367.5 & 103.74 \\
J2022+3842 & 20:22:21.6 & $+38$:42:14.8 & 1.95 & 429.1 & 48.58 \\
J2022+3845g & 20:22:11 & $+38$:45 & 1.76 & 487.5 & 1008.90 \\
J2024+3751g & 20:24:32 & $+37$:51 & 1.79 & 446.6 & 211.64 \\
J2030+3833g & 20:30:31 & $+38$:33 & 3.15 & 417.0 & 99999.9 $^a$ \\
J2030+3929g & 20:30:47 & $+39$:29 & 3.41 & 491.9 & 1718.42 \\
\hline
\end{tabular*}
    \begin{tablenotes}
        \footnotesize
         \item $^a$ Transient source with just few pulses, no period has been found yet \citep{zhx23}.
    \end{tablenotes}
\end{table*}

\begin{table}
\footnotesize
\caption{Timing parameters for PSR J2016+3711} \label{tab:tim_results}
\begin{tabular*}{\columnwidth}{lc}
\hline\hline
    Parameter & Value \\
    \hline
    MJD range & 59432--60196 \\
    Reference epoch (MJD) & 58871 \\
    RA, $\alpha$(J2000) & 20:16:09.14(6) \\
    Dec, $\delta$(J2000) & $+$37:11:10.4(1.1) \\
    Spin frequency, $f$ (Hz) & 19.68165408(6) \\
    First spin frequency derivative, 
    $\dot{f}\,$($10^{-11}\, {\rm Hz\,s}^{-1}$) & $-$2.8073(1) \\
    Second spin frequency derivative, 
    $\ddot{f}$ ($10^{-21}\, {\rm Hz\,s}^{-2}$) & 2.19(2) \\
    Dispersion measure, DM (pc cm$^{-3}$) & 428.0(1) \\
    Number of ToAs & 156 \\
    Residuals rms ($\mu$s) & 592.08 \\
    \hline
    \multicolumn{2}{c}{Derived Parameters} \\
    \hline
    Galactic longitude, $l$ (deg) & 74.9437 \\
    Galactic latitude $b$(deg) & $+$1.1142 \\
    Spin period, $P$ (ms) & 50.8087377(1) $^a$ \\
    First spin period derivative $\dot{P}\,(10^{-14}$ s s$^{-1}$) & 7.2472(4) \\
    Characteristic age, $\tau_{\rm c}$ (kyr) & 11.1 \\
    Surface magnetic field, $B$ ($10^{12}$ G) & 1.9 \\
    Spin-down luminosity, $\dot{E}$ ($10^{37}$ erg s$^{-1}$) & 2.2 \\
    Distance, $d$ (kpc) & 13.3 $^b$ \\
    \hline
\end{tabular*}
    \begin{tablenotes}
        \footnotesize
        \item $^a$ At the reference epoch (MJD) 58871.
         \item $^b$ Derived from the electron density model \citep{yao17}.
    \end{tablenotes} 
\end{table}

\subsection{The properties of the integrated radio pulse profile}
\begin{figure}
\centering
\includegraphics[width=\columnwidth]{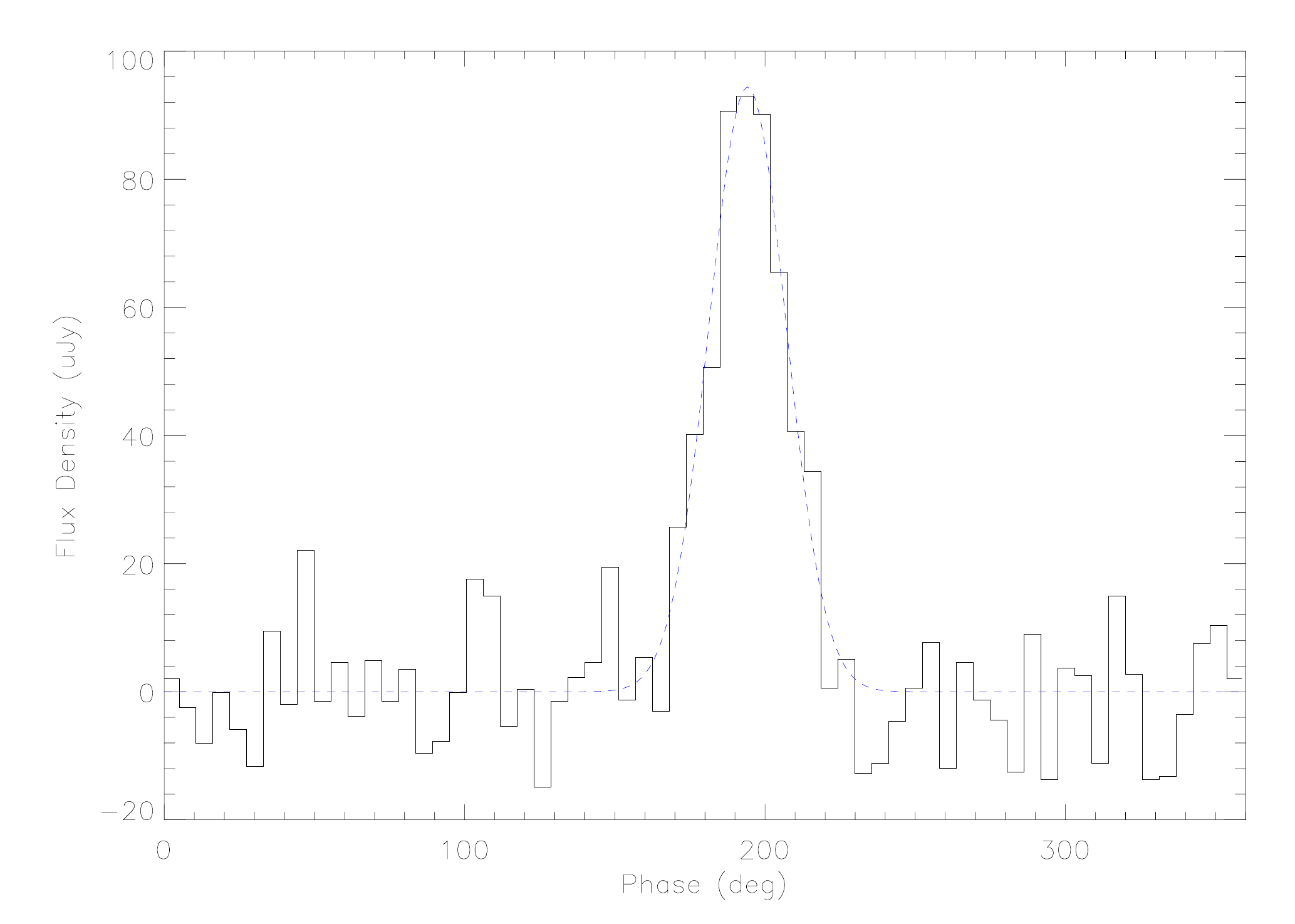}
\caption{Integrated pulse profile at 1.25\,GHz obtained from the FAST observation.
The blue dashed line represents the single component fitted with a Gaussian
curve.
}
\label{fig:integrate}
\end{figure}

The integrated radio pulse profiles are unique signatures that differ from pulsar to pulsar \citep[e.g.][]{lyn12}.
The integrated radio pulse of the pulsar in \snr, J2016$+$3711, can be well described by a single component with a Gaussian profile (Fig.\ \ref{fig:integrate}).
The width of the profile of a pulsar is usually measured at 10\% and 50\% of the peak flux density, as denoted by $W_{10}$ and $W_{50}$, respectively. 
Because the significance of the pulses detected here is low, we only estimate the value of $W_{50}$, which is about $28.1^\circ$, with an uncertainty of 
$\sigma_{50}=W_{50}/(\sqrt{2}\ln (2)S_{\rm S/N})
=2.7^\circ$, where $S_{\rm S/N}=10.8$ is the signal-to-noise ratio
of the radio pulse.
The 50\% duty cycle of the radio pulse is about
$(7.8\pm 0.8)$\%.

Also, we can estimate the effective pulse width $W_{\rm eff}$, 
which is defined as the width of a boxcar-like pulse with the same 
energy and amplitude of the real pulse \citep[e.g.][]{kra98}.
The value of $W_{\rm eff}$ is estimated to be $32.2^\circ\pm 1.7^\circ$,
which is equivalent to $4.5\pm 0.2$\,ms in units of time.

\section{Discussion}\label{discussion}

\subsection{The flux density of PSR J2016+3711}

The flux density of the radio pulses at 1.25\,GHz can be estimated as \citep[e.g.][]{dew85}
\begin{equation}
F_{\rm 1.25GHz}=
\frac{S_{\rm S/N}\beta}{(N_p \Delta \nu \, t_{\rm int})^{1/2}}
\left (\frac{W_{\rm eff}}{P-W_{\rm eff}}\right )^{1/2} \frac{T_{\rm sys}+T_{\rm sky}}{G},
\end{equation}
where $\beta \approx 1.5$ is a factor taking into account the losses
and system imperfections,
$N_p = 2$ the polarisation channels,
$\Delta \nu = 3.4\times 10^8$\,Hz the effective bandwidth of the data, 
$t_{\rm int} = 3470$\,s the on-source integrating time,
$P$
the radio pulsation period of the pulsar,
$T_{\rm sys} \approx 25$\,K \citep{jia19} the system temperature of the telescope, 
$T_{\rm sky} \sim 50$\,K \citep{kot20} the averaged background
temperature of the sky toward the pulsar at $\sim 1.25$\,GHz,
and $G \approx 16$\,K/Jy \citep{jia19} the gain of the telescope.
The flux density at 1.25\,GHz is estimated to be $15.5\pm 0.7\mu$Jy,
which indicates the faint nature of the pulsar's radio emission and explains why it has remained undetected in early surveys.
Notably, there are only 91 pulsars ($< 3\%$) in the ATNF catalog (v1.71) \citep{man05} with a flux density less than 16 $\mu$Jy at 1.4 GHz,
among which over 75\% were discovered by FAST \citep{han21}.

The flux densities of pulsars are known to have steep spectra \citep[e.g.][]{sie73},
with $F_\nu \propto \nu^{-\alpha}$, 
where $F_\nu$ is the flux density of the pulsar at frequency $\nu$,
and $\alpha$ the spectral index.
Here we crudely constrain the spectral index of PSR \pu,
based on the estimated flux density at 1.25\,GHz and the non-detection
at lower frequencies in early surveys.

The pulsar candidate in SNR CTB\,87 had been searched using the Low Frequency Array (LOFAR) at about 150\,MHz with a sensitivity of about 0.4\,mJy \citep{str19}, 
the Arecibo radio telescope at 430\,MHz with a sensitivity of 0.05\,mJy \citep{gor96},
and the 76\,m Lovell radio telescope at 606\,MHz with a sensitivity of 3\,mJy \citep{big96} and 0.16\,mJy \citep{lor98}, 
yet no radio pulse had been detected.
The $5\sigma$ values of the sensitivities are here used as the upper limits of the flux densities.

\begin{figure}
\centering
\includegraphics[width=\columnwidth]{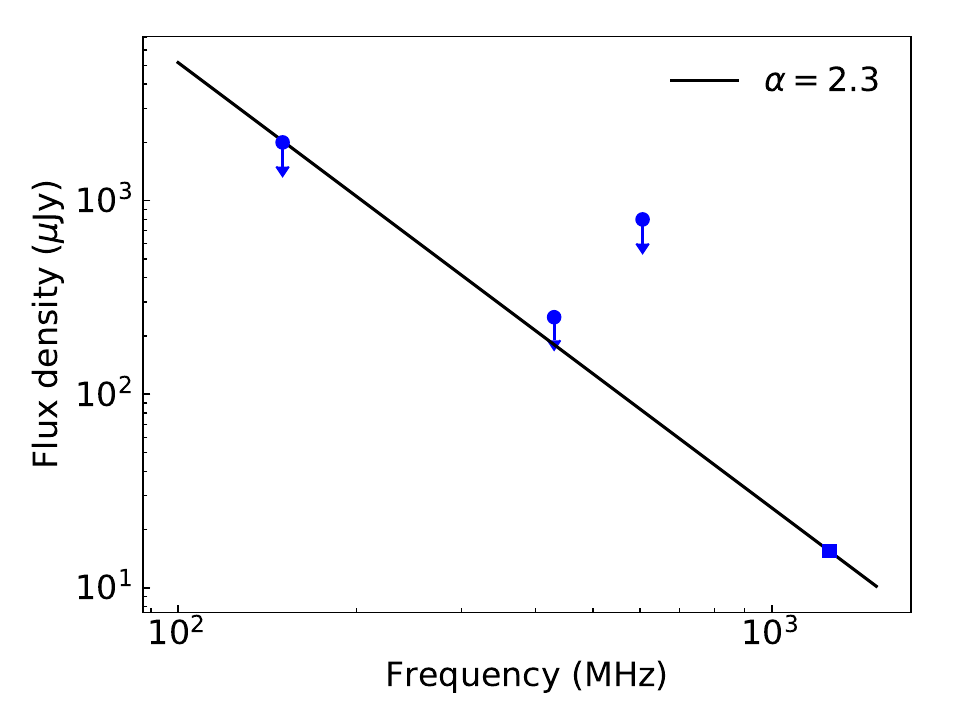}
\caption{Flux density of PSR \pu\ at 150\,MHz \citep{str19},
430\,MHz \citep{gor96}, 606\,MHz \citep{lor98} and 1250\,MHz. The 
$5\sigma$ values of the flux densities are shown as upper limits at
150\,MHz, 430\,MHz, and 606\,MHz.
The solid line represents the flux density distribution with spectral
index $\alpha$ equal to 2.3.
}
\label{fig:index}
\end{figure}

Fig.\ \ref{fig:index} shows the fitting of the flux density of PSR
\pu\ using $F_\nu = 15.5 (\nu/1.25\,{\rm GHz})^{-\alpha}\,\mu$Jy.
It is shown that the spectral index of the pulsar should not be larger than $\sim 2.3$.
It appears not to contradict the typical spectral indices of radio pulsars, which are in a range of
$\sim 1.4$--1.8 \citep[e.g.][]{lor95, tos98, mar00, bat13}.

It is noteworthy that the confinement is valid only if the spectral property of the pulsar can be well described by a single power law.
Actually, many young pulsars have broken power-law spectra,
and some even 
show a turn-over
at low frequencies
\citep[e.g.][]{bil20,jan18,mur17}. 
If there is a similar case for pulsar \pu, the spectral property of it would be more complex than derived above.

\subsection{The properties of PSR J2016+3711}
\label{pdot}
The observed period $P\approx50.81$\,ms and first period derivative $\dot{P}\approx7.25\times10^{-14}$\,s\,s$^{-1}$ are crucial to derive some basic physical properties of PSR~J2016+3711.
The characteristic age of the pulsar 
can be obtained through
$\tau_{\rm c} = P/[(n-1)\dot{P}] \approx 11.1$\,kyr for a dipole radiator with the braking index $n=3$.
Interestingly, the previous inferences using the empirical relation between the 
X-ray luminosity of PWNe and the characteristic age of pulsars, 
$\sim9.8
$\,kyr in \citet{mat13} and $\sim$11.5 kyr in \citet{guest20}
with a distance 6.1\,kpc adopted,
are very close to this estimate.

The strength of the equatorial surface dipole magnetic field,
can be inferred as
\citep{gae06}
$B = 3.2\times 10^{19}I_{45}^{1/2}R_{10}^{-3}(P\dot{P})^{1/2}$
$\approx 1.9\times 10^{12}I_{45}^{1/2}R_{10}^{-3}\tc ^{-1/2}$\,G,
where $I_{45}$ is the moment of inertia of the pulsar in units of 
$10^{45}$\,g\,cm$^{-2}$, and $R_{10}$ is the radius of the pulsar
in units of 10\,km.
Also, the spin-down luminosity of the pulsar is estimated to be $\dot{E}=4\pi^2 I_{45} \dot{P}/P^3 \approx 
2.2\times 10^{37}$ erg s$^{-1}$,
which can be used to explore the PWN scenario for the observed TeV gamma-ray emission \citep{veritas18}.

Using ${\rm DM}=428.0$\,pc\,${\rm cm}^{-3}$ and the electron density model \citep{yao17},
the distance to the pulsar is estimated as $\sim 13.3$\,kpc, which is about twice the estimate obtained from extinction-distance relation (6.1$\pm0.9$\,kpc) \citep{kot03}. 
This may be because
the line of sight will penetrate
the Orion spur, along which many objects (e.g. H\,{\small II} regions) should contribute significant DMs.
Furthermore,
we note that the distances to many pulsars estimated from the electron density model could
be overestimated 
for similar reasons
\cite[e.g.][]{yao17, han21}.

\subsection{Searching for a gamma-ray counterpart of the radio pulsation} 
\label{result:gamma}
Pulsars have long been suggested to emit gamma-rays \citep[e.g.][]{rom95}.
Actually, about 294 pulsars have been detected with pulsed gamma-ray emission to date \citep{fermi_psr23}.

To search for the possible gamma-ray pulsation,
we collected the \emph{Fermi}-LAT Pass 8 data toward a region that is centred at the pulsar with a radius of $1^\circ$,
from 2021-08-06 00:00:00 (UTC) to 2023-09-09 00:00:00 (UTC).
The data collected span over two years and cover the same MJD range as the timing analysis.
We analysed the data with the standard software {\small Fermitools} \footnote{\url{http://fermi.gsfc.nasa.gov/ssc/data/analysis/software}} version 2.0.8 released on 2021 January 20.
We only selected events with zenith angle $<90^\circ$ to filter out the background gamma-rays from the Earth's limb.
To optimise the signal-to-noise ratio, the analysis was restricted to the energy range 500 MeV--10 GeV.
We folded the \emph{Fermi}-LAT data with the ephemeris (parameter file) obtained from the radio timing results using the {\small TEMPO2} software \citep{edw06}
and checked the results using H-test.
Yet, no clear signature of pulsed profile was found.
We also performed the \emph{gtpsearch} routine of Fermi Science Tools software packages (v10r0p5) to search for pulsation frequencies near the radio pulsation frequency.
We adopted the spin frequency, the first and second spin frequency derivatives at 58871 MJD listed in Table \ref{tab:tim_results} for the search.
Still, there was no gamma-ray pulsation found.

Nonetheless, the existence of pulsed gamma-ray emission could not thus be excluded because of the limited statistics and the lack of a precisely determined pulsar ephemeris.
More follow-up radio observations spanning over years would be helpful to obtain a more accurate timing solution, which could then be used to fold the gamma-ray data and search for the pulsation.

\subsection{Orientation of the emitting regions}
The \emph{Chandra} X-ray observation of the CTB87 PWN displays a jets+torus structure \citep{mat13}. 
The projected elliptical shape of the torus suggests an inclination angle $\sim55^{\circ}$ for the equatorial wind. Similar to that in the G54.1$+$0.3 PWN, the torus is one-sidedly brightened and can be explained with a Doppler boosting of sub-relativistic downstream flow \citep{lu02}.
The jets, aligned with the pulsar spin axis, are at an angle $\sim55^{\circ}$ with the line of sight.

The radio pulse profile of PSR J2016$+$3711 is narrow ($W_{50}\approx28^{\circ}$, $W_{\rm eff}\approx32^{\circ}$), without broad wings, which is different from the pulse profile of PSR J1930$+$1852 in the G54.1$+$0.3 PWN with broad wings \citep{camilo02}.
This may indicate that either the pulsar radio beam starting near the magnetic polar cap is intrinsically narrow, or the line of sight sweeps just across a small segment of a broad beam. 
But, as a reference, the estimates of the beam radius from the relation obtained by \cite{kuz84} $\rho=5^{\circ}P^{-1/2}\sim22^{\circ}$
or the relation for the case of magnetic inclination angle close to $90^{\circ}$ \citep{lyne88} $\rho=6.5^{\circ}P^{-1/3}\sim17^{\circ}$ appear not in favor of the broad beam possibility.
The detection of the radio pulsation of PSR J2016$+$3711 implies no small magnetic
inclination angle (between the pulsar's spin, defined by the jets, and the rotating magnetic dipole axis, surrounded by the radio beam that sweeps past the Earth). 
If the gamma-ray pulsation is indeed undetectable, there could be two possibilities.
One could be that the pairs are not accelerated to the gamma-ray emitting energies.
The other could be that the direction angles of gamma-ray emission from the outer magnetosphere are offset from the magnetic dipole axis, avoiding the Earth.

\section{Summary}\label{summary}
We have performed a FAST observation to search for radio-pulsed signals toward two X-ray point-like sources in SNR \snr\ and proximate to SNR G22.7$-$0.02, respectively. 
We discovered radio pulses from the former with a significance of $\sim 10.8\sigma$, which is thus revealed to be a pulsar and named \pu.
This undoubtedly confirms the compact nature of the X-ray point source in SNR \snr.
This is the first pulsar discovered in SNRs using FAST thanks to its unprecedented sensitivities.
The integrated profile of the pulses can be well described by a single component, with a $W_{50}$ value of about 
$28.1^\circ$. 
Combining the estimated mean flux density at 1.25\,GHz
($\approx 15.5\mu$Jy) with the non-detection of the radio pulse at lower frequencies,
the spectral index of the pulsar is constrained to be $\lesssim 2.3$.
In the 28 follow-up timing observations with FAST,
we identified PSR \pu\ and measured a period of 50.8087377(1) ms, period derivative of $7.2 \times 10^{-14}$ s$\ps$, and DM of 428.0(1) pc cm$^{-3}$.
Additionally, the strength of the equatorial surface magnetic dipole magnetic field, characteristic
age and spin down luminosity of the pulsar are derived to be about 1.9 $\times 10^{12}$ G, 11.1 kyr, and 2.2 $\times 10^{37}$ erg s$^{-1}$, respectively.
Based on the radio timing results, we also searched for gamma-ray pulsation with the \emph{Fermi}-LAT data but found no pulsation.

\section*{Acknowledgements}

We are grateful to Zhichen Pan for the help with the timing analysis, and Jian Li for the discussion about the \emph{Fermi}-LAT gamma-ray data analysis.
WJZ thanks Xiao Zhang and Shijie Gao for helpful discussions.
YC thanks Jun Fang and Jin-Lin Han for helpful discussions on pulsar physics.
QCL acknowledges support from the Program A for 
Outstanding PhD candidate of Nanjing University.
YC acknowledges the support of
the National Key R\&D Program of China under grant
2017YFA0402600,
and the NSFC under grants 12173018 and 12121003. 
PW acknowledges the support from NSFC under grants 
11988101 and U2031117,
and the National SKA Program of China No. 2020SKA0120200.
PZ acknowledges the support from NSFC under grant 12273010.
YLY acknowledges support from
CAS ``Light of West China'' Program.
This work made use of data from FAST, a Chinese national mega-science 
facility built and operated by the National Astronomical Observatories, 
Chinese Academy of Sciences.

\section*{Data Availability}

The \emph{Fermi}-LAT data underlying this work are publicly available, and can be downloaded from 
\url{https://fermi.gsfc.nasa.gov/ssc/data/access/lat/}.
Over half of the FAST data we used are already in the public domain and can be accessed in accordance with FAST data policy.
The rest of undisclosed data can be obtained by contacting the corresponding author with a reasonable justification.



\bibliographystyle{mnras}
\bibliography{ref} 








\bsp	
\label{lastpage}
\end{document}